# Evolution of the dust in V4332 Sagittarii

Dipankar. P. K. Banerjee[1], Joseph A. Nuth III[2], Karl A. Misselt[3], Watson P. Varricatt[4], David Sand[5], N.M. Ashok[1], K. Y. L. Su[3], G.H. Marion[6], Massimo Marengo[7]

1 Astronomy and Astrophysics Division, Physical Research Laboratory, Ahmedabad, India 380009

2 NASA/GSFC , Mail Code: 690, Greenbelt , MD 20771, USA

3 Steward Observatory, University of Arizona, Tucson, AZ 85721-0065, USA

4 United Kingdom Infrared Telescope, 660 N. Aohoku Place, University Park Hilo, Hawaii - 96720, USA

5 Physics Department, Texas Tech University, Lubbock, TX , 79409, USA

6 University of Texas at Austin, 1 University Station C1400, Austin, TX, 78712-0259, USA

7 Physics and Astronomy Department, Iowa State University, Ames, IA 50011, USA

**Abstract**

An eruptive nova-like event took place in 1994 in the stellar-merger candidate V4332 Sgr. Following the eruption, dust consisting of refractory silicate rich dust grains containing a significant component of AlO bonding was formed sometime between 1998 and 2003. Observations using Spitzer between 2005 and 2009 show significant changes in the 10 micron silicate stretch feature. There is a deepening of the 10 micron silicate stretch as well as the development of a feature between about 13 and 20 microns consistent with a blend of the MgO and FeO stretching features and the O-Si-O bending mode of increasingly ordered silicate dust. Near-infrared observations show the presence of AlO and water vapor in the outflow in 2003, 2004 and 2005: the AlO has significantly decreased in spectra obtained in 2014 while the water vapor remains largely unchanged. An attempt is made to correlate these observations and understand the significance of these changes using DUSTY modeling. The observations appear consistent with the kinetically-controlled, condensation of highly under-oxidized SiO/AlO/Fe/Mg dust grains in the outflow followed by the continuous evolution of the initial condensate due to thermal annealing and oxidation of the dust via reaction with ambient O, OH and $H_2O$ in the expanding, cooling shell. Periodic monitoring of this dust shell over the mid-infrared spectral range could yield useful information on the evolution of under-oxidized silicate condensates exposed to hot water vapor in more conventional circumstellar environments.

# 1. Introduction

The nucleation and growth of oxide grains in circumstellar outflows is a poorly understood process even under the conditions found in winds around the ubiquitous Asymptotic Branch (AGB) Stars (Gail and Sedlmayr, 1998a,b; Posch et al. 1999, 2002; Zeidler et al. 2013, Gobrecht et al. 2015, Wittkowski et al. 2015). The composition of the condensate is difficult to predict empirically and the spectra of the resulting condensates do not conform to those of crystalline minerals except in the case of high mass-outflow-rate stars (Waters et al., 1996). While it was long assumed that the grains would be primarily silicates (Nuth & Donn 1982, Gail & Sedlmayr 1986) and the observed spectral properties of such grains are certainly consistent with the presence of amorphous silicate (Habing 1996, Danchi et al. 1994), attempts to model silicate condensation in such systems generally failed (Gail and Sedlmayr, 1998b) due to the calculated instability of silicate dust at the temperatures observed in the dust shell. This problem has recently been solved (Nuth and Ferguson, 2006) based on new measurements of the vapor pressure of SiO (Ferguson and Nuth, 2008) that differ from those previously used to model silicate nucleation and growth (Gail and Sedlmayr, 1998a) by nearly 5 orders of magnitude at 1000K.

A less appreciated problem in understanding the composition of astrophysical condensates is the chemical kinetic inhibition that retards formation of the most thermodynamically stable mineral grains from the outflowing gases. Most refractory species in stellar outflows consist of simple metals or oxides such as Fe, Mg, Al, SiO or AlO (Tsuji, 1973) and the formation of grains requires such vapors to become supersaturated. However, once solid surfaces do form refractory vapors are rapidly depleted. The result is that molecular SiO or AlO condenses but does not get a chance to react with O, OH or $H_2O$ to form the fully oxidized solids that are typically encountered in the terrestrial environment (Nuth, Rietmeijer and Hill, 2002). Measurements of the spectra of rapidly condensed grains formed from complex vapor mixtures including Fe, Mg, SiO and AlO in a hydrogen rich atmosphere (Nuth, 1996) demonstrated that spectral features associated with the AlO or SiO stretching features were not initially found in proportion to the concentration of AlO or SiO in the solids. Such spectra did become more "rational" following hydration or thermal annealing of the initial condensates once the grains had the chance to become more completely oxidized and closer to equilibrium.

Opportunities to observe grain processing in astrophysical environments are rare; as typical AGB winds appear as quasi-steady state flows and the spectrum of their dust is either an average of the initial condensate processed to some ill-defined degree for optically thin shells, or emanates from the outer layers of optically thick shells (Forest et al., 1978). Better opportunities to follow the evolution of fresh condensates are available for materials produced in variable or flare stars (Stencel et al., 1990) where dust formed in a single event can be followed over many years. The eruptive variable V4332 Sgr provides an excellent opportunity for such a study (Banerjee et al., 2007).

V4332 Sgr erupted in 1994 in what was initially considered a nova-like outburst (Martini et al. 1999). However, its subsequent post-outburst evolution to a cool spectral type indicated that this was not a classical nova eruption. The light curve of the object (Martini et al. 1999) showed a slow rise to a maximum visual magnitude of ~ 8.5 which was followed by a fast decline with a decay time for 2 and 3 magnitudes of 8±1 and 12±1 days, respectively. The exact nature of V4332 Sgr is of considerable interest as it, along with such objects like V838 Mon and M31 RV (a red-variable which erupted in M31 in 1988), may form a new class of eruptive objects (red novae) which originate from stellar mergers (e.g., Munari et al. 2002; Banerjee and Ashok 2002, Bond et al. 2003). The post-merger evolution of such objects is a field of considerable current interest (Kaminski et al. 2013, 2015 and references therein; Chesneau et al 2014, Loebman et al 2015). In May 1998 V4332 Sgr was observed by 2MASS and its JHK colors showed no evidence of infrared excess associated with dust. At this stage it had the SED of a M type star with T = 3250K. Observations in June 2003 showed it had developed an IR excess due to dust (see Figure 2 in Banerjee et al., 2003). Thus dust formed around the object in a shell or disk between 1998 and 2003. This dust is now 12 to 17 years old and provides a rare and valuable opportunity to study the generic evolution of fresh condensates in dust. The properties and mineralogy of this dust shell was studied extensively in Banerjee et al (2007) using Spitzer data. V4332 Sgr in general shows a rich emission line spectrum both in the optical and NIR (Tylenda et al 2005, Banerjee and Ashok, 2004); in addition to emission from the fundamental band of 12CO at 4.67 micron along with water ice at 3.05 micron (Banerjee et al. 2004)

Dust formed in the outflow of V4332 Sgr could not be modeled using amorphous silicate grains alone: a significant component of the dust spectrum was shown to be consistent with the presence of amorphous Alumina as an identifiable component of mixed amorphous oxide dust (Banerjee et al 2007). One possibility to explain the observed spectrum was that mixed metal oxide grains had formed but that the high oxygen affinity of aluminum had rapidly resulted in the total oxidation of aluminum and the reduction of at least a portion of the SiO to metal (Banerjee et al., 2007) as had been previously predicted (Nuth and Hecht, 1990). If this scenario was correct then it should be possible to follow the subsequent oxidation of these grains by observing the spectral evolution of the dust over time.

## 2. Observations:

Four ground-based spectra were obtained in the NIR region between 2003 and 2014; three between April 2003 and July 2005 with the 3.8m UK Infrared telescope (UKIRT) and one in September 2014 using the 3m IRTF. Each of the 3 UKIRT spectra were obtained using the UKIRT 1-5 micron Imager Spectrometer (UIST), the HK grism, the 4 pixel wide slit which gives a resolution of 500 and an exposure time of 720s. Data reduction was done along similar lines as described earlier in Banerjee et al (2003). The IRTF spectrum was obtained using SpeX (Rayner et al. 2003) in the cross-dispersed mode using the 0.3"x15" slit (R = 2000) and a total exposure time of 1200s. The SpeX data were reduced and calibrated using the Spextool software (Cushing et al. 2004), and corrections for telluric absorption were performed using the IDL tool xtellcor (Vacca et al. 2003) . Spitzer observations were carried out under program S50106 in cycle GO5 in the IRS stare mode using the Short Low 2, Short Low 1, Short High and Long High modules. The log of the observations is given below in Table 1. The 2003 and 2004 UKIRT spectra have been presented elsewhere earlier (Banerjee et al. 2003, Banerjee et al 2004) but since they are important for the discussion they are included here.

**Table 1: Log of the observations**

| Date | Instrument/Observatory | Wavelength range |
| --- | --- | --- |
| 23 April 2003 | UIST on UKIRT | 1.4 – 2.5 micron |
| 15 April 2004 | UIST on UKIRT | 1.4 – 2.5 micron |
| 10 July 2005 | UIST on UKIRT | 1.4 – 2.5 micron |
| 7 May 2009 | IRS on Spitzer | 5 to 35 micron |
| 23 Sept 2014 | SPeX on IRTF | 0.8 to 2.5 micron |

## 3. Results:

We present the 5 to 35 micron Spitzer spectra in Figures 1 and 2. The Spitzer 2005 October 15 and 2006 November 2 spectra have been presented in Banerjee et al (2007) earlier but are included here to show the temporal evolution of the dust features. In Figure 3 we show the evolution of the near-IR (1 to 2.5 micron) ground based spectra between 2003 to 2014. We discuss these spectra below.

In comparing the Spitzer dust spectrum of V4332 Sgr obtained in 2005 with that obtained in 2009 (Figs 1 and 2), it is easy to see that there has been some evolution in the 8 to 20 micron region. A convenient way to see the change in this region is by subtracting the underlying continuum. To set the endpoints for the continuum subtraction, we have chosen the MIPS 24 micron point on the redside and on the blue side the first data point of the SL1 module at 7.5 microns. The choice of the 7.5 micron end-point is for the following reason. Data shortward of 7.5 micron is covered by the SL2 module (which covers 5.0 to 7.5 micron) and because there is a small wavelength gap between the SL1 (which covers 7.5 to 14 micron) and SL2 modules, joining their ends by matching the flux at a common wavelength is difficult. Hence we leave out the 5.0 to 7.5 micron data from the continuum subtraction.

The continuum subtracted spectrum is shown in Figure 4. For simplicity a linear continuum was used; a non-linear continuum approximated by a polynomial (cubic) fit was tried and not found to give results significantly different from that shown in Figure 4. On the figure we mark the expected positions of the silicate feature using a nominal value at 10 micron and the minimum of the porous alumina absorption feature at 11.3 micron based on the Begemann et al. (1997) optical constants. The major changes that can be seen, using Figures 1 and 4 in combination, are as follows. First, the relative depth of the 10 micron absorption has increased significantly. Most of the deepening is at the position of the silicate component suggesting it has increased in content. More SiO may have condensed or there may be more SiO bonds – possibly due to the oxidation of the initial SiO in the dust. The red wing of the AlO component has eroded and drops steeply between 11.3 to 13 micron. These qualtitative changes suggesting an increasing silicate and decreasing alumina content between 2005 to 2009 are quantitatively supported through radiative transfer modeling done in later sections. A second difference is the development of a broad bump starting from 13 microns and extending up to ~19 to 20 microns (this bump is also seen in Figure 1). A third change seen from Figure 1, though not central to the paper, is that the water ice feature at 6 microns has remained virtually unchanged between the observations taken in 2005 and 2009.

To understand these changes we have modeled the V4332 Sgr spectrum using the radiation transfer code DUSTY (Ivezic et al. 1999). We first explored the extent to which annealing of the silicate component of the dust could account for the observed changes in the spectrum. Before proceeding further, we review the results of the DUSTY modeling of the 2005 Spitzer spectrum in Banerjee et al. (2007). In that work it was seen that the observed SED, and specifically the extremely broad 10 micron feature normally associated with the presence of amorphous Mg-Fe silicate grains, could not be modeled by silicate dust only (see Figure 1, lower panel in Banerjee et al. 2007) . An additional component of alumina dust had necessarily to be introduced to reproduce the width of the feature. The inclusion of alumina was prompted by the presence of a feature at ~11 micron often attributed to amorphous alumina (Speck et al. 2000; Posch et al. 2002, DePew et al. 2006). It is worth mentioning that in AGB stars, the amorphous 11μm $Al_2O_3$ feature is often seen in combination with a narrow, crystalline $Al_2O_3$/$MgAl_2O_4$ feature at 13μm (Posch et al. 1999, Zeidler et al. 2013) and an (Mg,Fe)O feature at 19.5μm (Posch et al 2002), but this does not seem to be the case in V4332 Sagittarii. We are also not aware of any other object like V4332 Sgr which shows the ~ 11 micron Alumina feature in absorption; all reported detections of this feature, which we have come across, show it to be in emission. But it would be worthwhile to seek for analogous Alumina signatures in similar recently discovered stellar-merger objects like V1309 Sco or V838 Mon (Kaminski et al. 2013, 2015). An attempt in this direction, to record their mid-IR spectra, using airborne platforms is already underway.

While the effects of annealing on alumina dust are unknown, the effects of annealing of silicate dust on the SED has been studied by Hallenbeck et al (2000; hereafter H2000) . We thus construct a series of model DUSTY spectra, discussed shortly, in which the silicate component has been allowed to thermally evolve as the result of annealing based on the H2000 results. To appreciate these models, we briefly summarize the essential results from H2000.

Laboratory studies were made in H2000 of the evolution of a magnesium silicate smoke from an amorphous condensate to a crystalline mineral by annealing in vacuum at temperatures ranging from 1000 to 1200 K. This provided a basis for developing a silicate evolution index (SEI) which could be used to predict the emergent IR spectrum of freshly condensed silicate dust as it anneals in a circumstellar environment. The initial smoke sample (Nuth, Hallenbeck, & Rietmeijer 1999) was a mixture of amorphous magnesium silicate grains that span a broad range of Mg/Si ratios peaking at compositions approximating the smectite ($Mg_6Si_8O_{22}$) and serpentine ($Mg_3Si_2O_7$) dehydroxylates. The IR spectra of the annealed smokes were recorded as a function of the degree of thermal evolution of the samples and the optical constants derived at different stages of the evolution (i.e at different SEI values). Worth mentioning was the finding of a natural pause or "stall" in the spectral evolution of the sample, midway between the initially chaotic condensate and the more ordered glass. Thereafter, individual features sharpened as the sample became more ordered. After removing the contribution of pure silica from the recorded IR spectra, H2000 present the optical constants for pure magnesium silicate dust at different stages of thermal processing (see their Tables 1 and 2). Using the Draine and Lee (1984) "astronomical silicate" as a template, they also provide optical constants for a nonsilicate oxide component to the dust that might be found in oxygen-rich circumstellar shells. This component is derived by subtracting the spectrum of unannealed magnesium silicate smoke from the astronomical silicate of Draine & Lee (1984) - greater details are provided in H2000.

3.1 DUSTY modeling:

In our DUSTY modeling we have therefore used three components. The first is a magnesium silicate component at different evolutionary stages, the second component comprises of astronomical silicate minus magnesium silicate (e.g., other oxides) which we will call the non-Mg silicate component and whose optical constants are listed in Table 3 of H2000. The third component is amorphous alumina whose optical constants are provided in Begemann et al. (1997). Since not much is known about the annealing behavior of the non-Mg silicate nor the amorphous alumina components, we have assumed that their spectra do not change with annealing. It should be noted that DUSTY requires the optical constants to be listed as real and imaginary parts (n,k) of the complex refractive index $m = n + ik$. However, the optical constants listed in the various tables of H2000 are given in terms of Re ($\varepsilon$-1) and Im ($\varepsilon$) where $\varepsilon$ is the complex dielectric function. These may be converted to the (n,k) format by using $m = \varepsilon^{1/2}$.

The results of our simulations are shown in the six sub-panels of Figure 5. These sample, in a reasonably equi-spaced manner, the entire evolution of the silicate annealing process as given in H2000 from an unannealed phase through the pre-stall, stall and post-stall phases. By varying the parameters and specially the dust composition, it was first determined at which stage we could get the best model fits ; this was seen to be during the post-stall phases. Panels e and f represent reasonably good model fits. The fractional abundances by number, of the grains of the different dust components (MgSil; non-MgSil and Alumina) was then fixed at the values found in the best fits and this was retained for the fits in all other panels. In this manner the effects of annealing could be visualized since the most significant factor that was changed between fits was the optical constants of the MgSil corresponding to different phases of the SEI. Between fits, we also changed the optical depth factor ($\tau$) that is used to adjust the optical depth of the dust at a chosen wavelength – but variations in $\tau$ were found to be small. The parameters of the different fits are listed in Table 2. All other parameters which characterize the V4332 Sgr system were retained from Banerjee et al (2007; see Table 1 therein) and held fixed for all fits. These are as follows: stellar luminosity of $10^4$ $L_{sun}$, stellar temperature of 3250 K, an outer-to-inner dust shell radius of 1000, a shell density distribution that varies as $r^{-2}$, a dust temperature at the inner radius of the dust shell of 1750 K and an MRN grain size distribution (Mathis et al. 1977).

The inferences that can be drawn from Figure 5 are as follows. First, it is seen from panel 5(a) that a pure silicate dust composition does not provide a reasonable fit to the observed SED. It is necessary to include a substantial amount of alumina; a conclusion consistent with that reached by Banerjee et al. (2007) Second, it is seen that poor fits are obtained with unannealed and less annealed (SEI less than 1) MgSil samples (panels a,b,c,d). The fit is not only poor over the body of the 10 micron profile but the minimum of the absorption dip of the model fits occur blueward of the observed minimum. As stall is crossed, the fit improves and further the minima of model and observed data begin to match well. This indicates that the dust appears to be well annealed in V4332 Sgr even as far back as in 2005. However, even the better fits of panels e and f, show deviations from the observed data – especially for wavelengths longer than 18 micron (or log$\lambda$=1.25). This is true for the 2005 and 2009 data, however for the 2009 data, the deviations are slightly worse as shown in Figure 6 which is discussed shortly.
In short, we find that the changes seen between the 2005 and 2009 spectra cannot be explained in totality by two different sets of the H2000 data with progressively larger SEI values i.e an increase in the degree of annealing with time. In general the deviations of the best fits H2000 models from the observed data could be caused by a variety of factors such as not considering the effects of annealing of the Non MgSil and alumina components or taking into account the effects of oxidation or a combination of both these factors.

In an effort to get better fits to the 2005 and 2009 spectra and the change between them, we tried other material compositions in the DUSTY modeling. We find that a dust composed of the pyroxene glassy silicate of Jaeger et al (1994) when combined with alumina reproduces the observed data quite well. The glassy silicates of Jaeger et al (1994) were prepared in the laboratory with the aim of replicating a mean cosmic composition with respect to the four most abundant metals in the cation content. The silicate spectra generated by using these glassy silicates were found to match the spectra of star-forming regions and YSOs quite well. Figure 6 (panels b and a respectively) show DUSTY models of such fits for both the 2005 and 2009 data. One of the better post-stall fits of H2000 from Figure 5(f) is shown for comparison in panel 6(d) for the 2005 data and repeated again for the 2009 data in panel 6(e). The parameters of all the fits are listed in Table 2. For the 2005 spectrum, we have included the MIPS 70 and 160 micron flux values of 1.07 and 0.12 Jy respectively for which we have data (Banerjee et al. 2007). It may be seen that the glassy silicate models fit both the 2005 and 2009 data quite well. There is a compositional difference between the two epochs with the silicate content increasing with time by ~ 7% accompanied by a decrease in the alumina content by the same amount as the data in Table 2 indicates. The amount of change is small but it could be an important indication. The robust way to conclude whether this change persists is through further observations which should show whether the trend continues. For the moment, the observed behavior of an increasing silicate and decreasing alumina composition would be consistent with the chaotic silicate hypothesis of Nuth and Hecht (1990) which is thus discussed in the coming section.

## Table 2: Parameters of DUSTY model fits

| Figure Number | Composition | Opt. depth $\tau_{9.8}$ |
|---|---|---|
| 5(a) blue curve | 0.27 (MgSil[1]) 0.10 (Non MgSil) 0.63 (Alumina) | 46.5 |
| 5(a) green curve | 0.50 (MgSil) 0.50 (Non MgSil) 0.00 (Alumina) | 48.5 |
| 5(a) red curve | 0.00 (MgSil) 0.00 (Non MgSil) 1.00 (Alumina) | 24.5 |
| 5(b) | 0.27 (MgSil) 0.1 (Non MgSil) 0.63 (Alumina) | 45.0 |
| 5(c) | 0.27 (MgSil) 0.1 (Non MgSil) 0.63 (Alumina) | 45.0 |
| 5(d) | 0.27 (MgSil) 0.1 (Non MgSil) 0.63 (Alumina) | 46.5 |
| 5(e) | 0.27 (MgSil) 0.1 (Non MgSil) 0.63 (Alumina) | 48.0 |
| 5(f) | 0.28 (MgSil) 0.1 (Non MgSil) 0.62 (Alumina) | 49.0 |
| 6(a) | 0.40 (glsil) 0.60 (Alumina) | 45.0 |
| 6(b) | 0.47 (glsil) 0.53 (Alumina) | 43.5 |
| 6(c) | 0.28 (MgSil) 0.1 (Non MgSil) 0.62 (Alumina) | 49.0 |
| 6(d) | 0.28 (MgSil) 0.1 (Non MgSil) 0.62 (Alumina) | 49.0 |

(1) Optical constants of MgSil, Non MgSil from H2000; glassy silicate from Jaeger et al. (1994); Alumina from Begemann et al. 1997

## 4. DISCUSSION

Many of the changes observed in the spectrum of V4332 Sgr appear to be consistent with the predictions of Nuth and Hecht (1990) provided that the major change in dust chemistry that produced the observed spectral changes is due to oxidation. If we assume that the initial condensate observed in 2005 was both hot as well as under oxidized, then the initial chemical equilibration within the grains would lead to the formation of AlO bonds at the expense of all other metallic species. Since SiO has the second highest affinity for oxygen after aluminum, and since aluminum is generally only 10% of the abundance of Si, some SiO bonds would naturally be present in the initial grain population. However, if the SiO is not fully oxidized then it is highly unlikely that any other metal oxides (MgO or FeO) would be present in these grains.

Once the refractory species had all been depleted from the vapor phase, the solid grains would continue to evolve as they slowly react with the hot gas in which they are embedded. As oxygen from ambient O, OH or $H_2O$ diffuses into the dust grains, silicon metal and silicon monoxide will first oxidize to the $(SiO_4)^{-4}$ anion that is the basis for all silicate glasses and minerals. Once the majority of the silicon in the grains was fully oxidized, other metals such as the Fe or Mg would become oxidized as well.

As the silicon oxidizes to $(SiO_4)^{-4}$ the SiO stretch grows to overwhelm the AlO absorption that was at its maximum in the original condensate. It is even possible that individual aluminum atoms (as aluminum oxide) become incorporated into the growing amorphous silicate lattice. The result is a deeper and narrower absorption feature near 10 microns (as seen in Figures 1 and 4). This should be accompanied by a gradual disappearance of the red wing due to the AlO stretch. In other words, we would expect the pure alumina content to decrease as the dust expands and cools relative to the growing silicate spectral signature and some fraction of the initial AlO is incorporated into the silicate itself. As discussed in the preceding section, such a trend is seen from our modeling though the alumina decrease is small. One possible reason for this is that whatever is lost due to oxidation is replenished by fresh alumina being produced from condensation of the gas phase AlO. As can be seen from Figure 3, there was a drastic decrease in the gas phase AlO between 2005 and 2014. If a large fraction of this AlO decrease were to be channeled into production of fresh alumina, then the relative proportion of alumina with respect to the silicate content would not be expected to drop sharply. For this to happen, it needs to be ascertained that gas phase AlO is indeed linked to alumina production. There is evidence in that direction from the study of Banerjee et al (2012) wherein a search was made for fresh detections of AlO in stars with a view to better understand the characteristics of this radical. The majority of the AlO detections, totaling 13 out of 17 targets studied, were in AGB stars all of whose IRAS spectra either showed a broader than usual 10 $\mu$m silicate feature suggestive of the presence of alumina or a clearly discernible 11 $\mu$m alumina feature on its red wing. In the context of the decreasing gas-phase AlO content, another comment may be made. From Figure 3 it is obvious that AlO remained in the vapor for several years after dust formation (2003 – 2005) although the signature of AlO, while still visible, had decreased markedly by 2014. In comparison, the relative intensity of the water signature has remained more or less constant throughout the entire period of these observations as would be expected if circumstellar water vapor were either present in great excess of all oxidizable, refractory elements or if the oxidation of such elements occurred at a very slow rate. Based on the decrease in the AlO signature we believe that the first explanation is more probable.

Apart from the changes to the silicate and alumina related features, oxidation of MgO and FeO would naturally result in a broad absorption feature between 14 – 20 microns depending on the diversity of the chemical environments within the grains. Such a feature is again seen in Figures 1 and 4. As annealing or oxidation proceeds and the amorphous silicate progresses from condensed SiO molecules to more uniform and highly polymerized $(SiO_4)^{-4}$ dust grains the O-Si-O bending mode just beyond 20 microns should become more prominent in the dust spectrum. This is not yet obvious in the spectrum shown in Figure 1 and may never develop due to the relatively low temperatures of the outermost dust grains. The DUSTY code that was used to fit the SED in Banerjee et al (2007) and in the present modeling generates a radial profile of the dust temperature. These results show that if the T(dust) = 1750K at the innermost radius $r_0$ (the model value used by us) then it falls to 1000K at $\sim 2r_o$ and to about 500K at $\sim 6r_o$. While pure thermal annealing is important within about $2r_0$, oxidation should be the dominant driver of the spectral evolution outside this radius. The outermost material is absorbing against a background of hotter dust grains. At $1000r_o$ it falls to about 40K ($1000\ r_o$ is chosen as the dust shell thickness) and is certainly cool enough to explain the unchanging nature of the 6 micron ice band.

What might we expect to see as (if) the dust grains around V4332 Sgr continue to evolve? Based on the changes so far observed in this system, we can predict that the broad feature observed between 14 – 20 microns will deepen and begin to narrow as the grains become more homogeneous and provide a narrower range of chemical environments housing Mg and Fe cations. The feature should also shift to longer wavelengths as the O-Si-O bending mode becomes a more prominent part of this blended feature and the short wavelength side of the feature narrows due to the increasing homogeneity of the environment surrounding the MgO and FeO stretching modes.

One caveat to this prediction is based on the grains remaining sufficiently hot so that they continue to react both internally and with the surrounding gas. Oxidation of the individual grains could provide some of the activation energy required to rearrange chemical bonds within the dust (see Ishizuka et al., 2015). However, the activation energy for grain annealing is a very steep function of temperature (Hallenbeck et al., 1998, 2000) and requires temperatures in excess of 900 – 1000K to obtain significant changes to the internal grain structure on timescales of only years. Therefore it is possible that grain cooling may quench the progression towards the formation of crystalline minerals well before such materials can form. To a large extent, it will depend on what kind of unusual object is left at the center of V4332 Sgr, now that a merger has taken place, and the nature and amount of radiation it provides to keep the grains hot. At present the central remnant has the spectral characteristics of a M6 giant.

The detailed study of the dust features in the 10 micron infrared spectra of oxygen-rich evolved stars by Speck et al (2000 ) showed that while the "classic" silicate feature is essentially identical for both AGB stars and red supergiants, the "broad" features – associated with he presence of Alumina in addition to silicates - observed for these two stellar types are quite different. They suggested that the dust in these two environments followed different evolutionary paths, with the dust around Mira stars, whose broad feature spectra can be fit by a combination of alumina ($Al_2O_3$) and magnesium silicate, progressing from this composition to dust dominated by magnesium silicate only, while the dust around supergiants, whose broad feature can be fit by a combination of Ca-Al-rich silicate and Al2O3, progresses from this initial composition to one eventually also dominated by magnesium silicate. Thus there is a similar trend seen in the present work and that of Speck et al. (2000). But the reader is directed to this latter work for the discussion on the complexity of the situation in AGB stellar spectra.

In the 21 years since its outburst in 1994, V4332 Sgr has shown interesting developments in its IR evolution. Here we have shown through Spitzer observations that the 10 micron silicate feature and the regions extending redward from it up to 20 micron has evolved significantly between 2005 and 2009. The change is clearly seen although understanding the reasons for the change do not appear straightforward. It is always possible that a better model, than that presented here, could be constructed which considers the effects caused by a larger range of variations in the input physical parameters for e.g. the grain size distribution and geometry. But on the whole V4332 Sgr provides a good opportunity for observations with future space missions to monitor further evolution brought about by the oxidation and annealing of dust grains in an O-rich environment.

## 4. Summary

We have presented data, using Spitzer and ground-based observations obtained between 2005 and 2009, to document the evolution of the freshly condensed dust that formed in the eruptive variable V4332 Sgr after its 1994 outburst. The evolution in the spectra has been analysed using the radiative transfer code DUSTY. Spectra obtained in 2005 showed the presence of dust with a significant component of AlO bonds. Spectra obtained in 2009 show that the spectrum of the dust around this star has evolved in response to oxidation by the hot gas in the outflow with the growth of the SiO stretching feature as well as the growth of stretching features due to both MgO and FeO between 13 – 20 microns. As these silicate and metal oxide features grew, the red wing associated with the AlO stretch is seen to have decreased in strength. If this trend has continued to the present we would expect to see further narrowing in the 10 micron feature as well as continued compaction of the longer wavelength feature. As the dust evolution appears to be ongoing it is important to document the present changes and observations to monitor further changes.

## 5. Acknowledgements:

Research at the Physical Research Laboratory is supported by the Department of Space, Government of India. DS is a Visiting Astronomer at the Infrared Telescope Facility, which is operated by the University of Hawaii under contract NNH14CK55B with the National Aeronautics and Space Administration. He would like to thank Michael Cushing for his efforts in keeping the Spextool reduction code up to date. We are grateful to an anonymous referee for helpful comments that improved the paper and deeply appreciative for the critical, yet favorable, evaluation of this work.

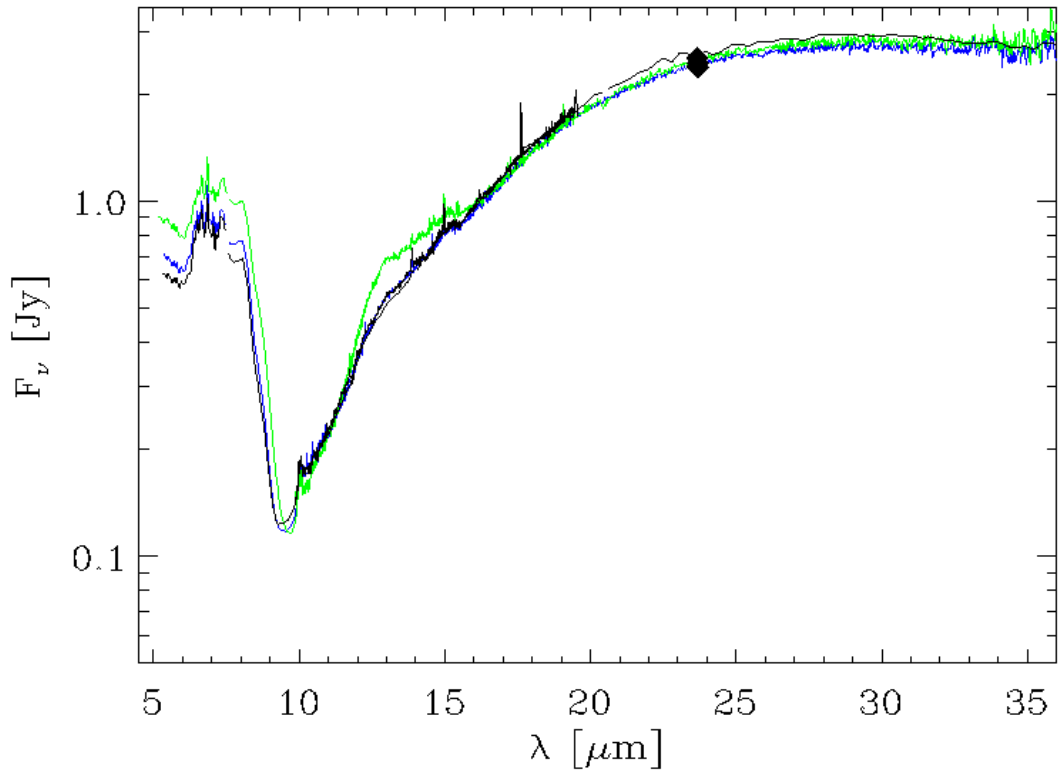

Figure 1: The Spitzer spectrum of V4332 Sgr of October 2005 (black), November 2006 (blue) and May 2009 (green). The observed MIPS fluxes at 24 microns is shown by a black filled dot which have values of 2.526, 2.388 and 2.451 Jy at the 3 epochs respectively.

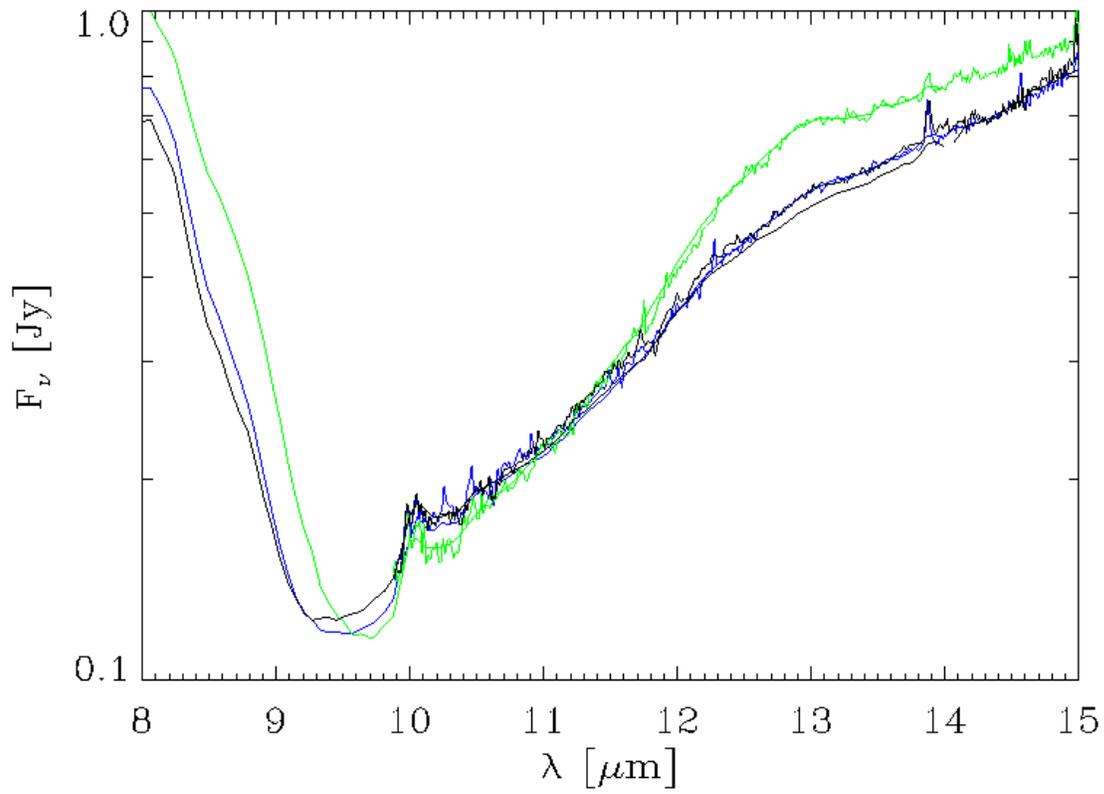

Figure 2: A magnified view of the Spitzer spectrum of V4332 Sgr around the 10 micron silicate feature to show the temporal evolution. Color code for the spectra: October 2005 (black), November 2006 (blue) and May 2009 (green).

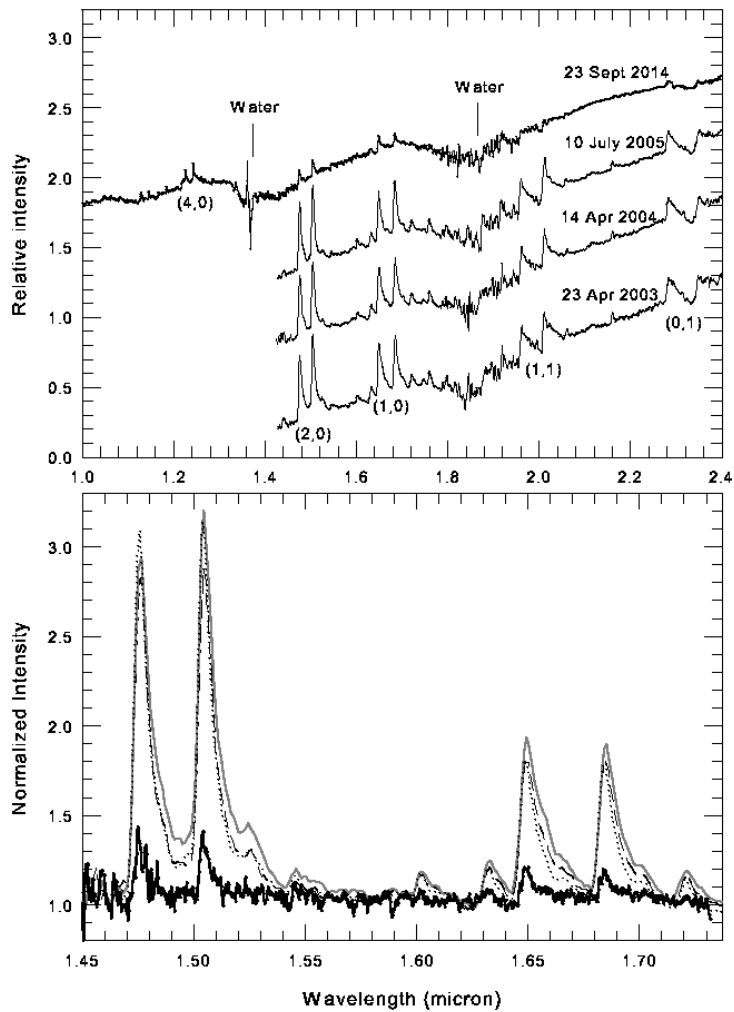

Fig3: The top panel shows the time evolution of the NIR spectra (1 to 2.5 micron) with the strongest AlO A-X bands marked. The bandheads for the (1, 0) band occur at 1.6837 and 1.6480 micron, for (1, 1) at 2.0106 and 1.9600 micron, for (2, 0) at 1.5035 and 1.4749 micron and for (4, 0) at 1.2425 and 1.2256 micron. Each spectrum has been normalized to unity at the K band center (2.2 microns) and then offset for clarity. The bottom panel shows a magnified view of some of the AlO bands, in which the continuum of each spectrum has been normalized to unity. It illustrates the sharp decrease between 2005 to 2014 of the strengths of the AlO bands. The spectra of 2003, 2004, 2005 and 2013 are shown by the gray, dashed, dotted and dark continuous lines respectively. The top panel also shows the presence of water vapor in the shell between 2005 and 2014 as evidenced from the turning down of the edges of the spectra in the 1.35 to 1.45 micron region and 1.75 to 1.9 micron region. This is a typical sign of the presence of water in the spectra of cool, late type stars.

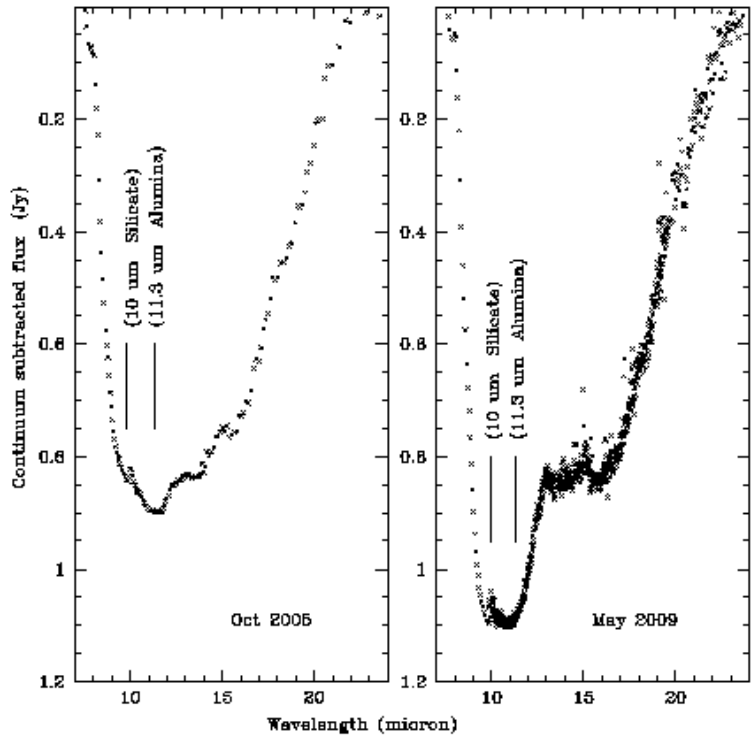

Figure 4: The spectral region around the 10 micron silicate feature, after a continuum subtraction, to facilitate seeing the changes in the spectra between 2005 and 2009. The nominally expected positions of the silicate and alumina absorption minima are marked at 10 and 11.3 micron. Detailed discussion is given in section 3 (paragraph 3).

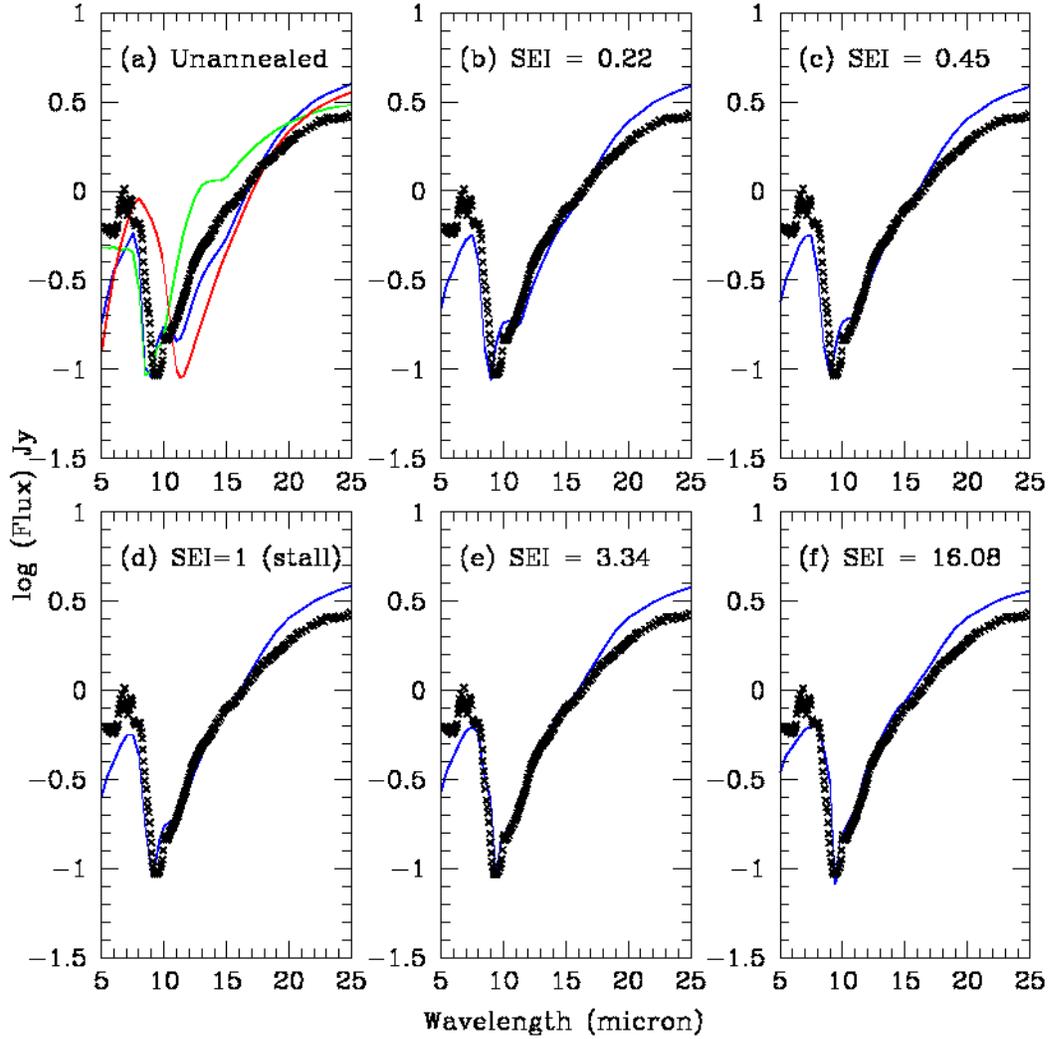

Figure 5: An analysis of the effect of annealing on the observed SED using DUSTY modeling. Increasing effects of annealing of the silicate component, as characterized by an increasing silicate evolution index (SEI), are shown in panels (a) to (f) by using different sets of optical constants for the different SEI's from Hallenbeck et al. (2000). In each panel the black crosses are the observed Spitzer data for October 2005. The blue lines in all panels are DUSTY fits using a composition of silicates and porous alumina dust. In panel (a) the green curve is for a pure silicate composition while the red curve is for a pure alumina composition. The chemical composition for each curve is given in Table 2 and greater details are presented in section 3.1

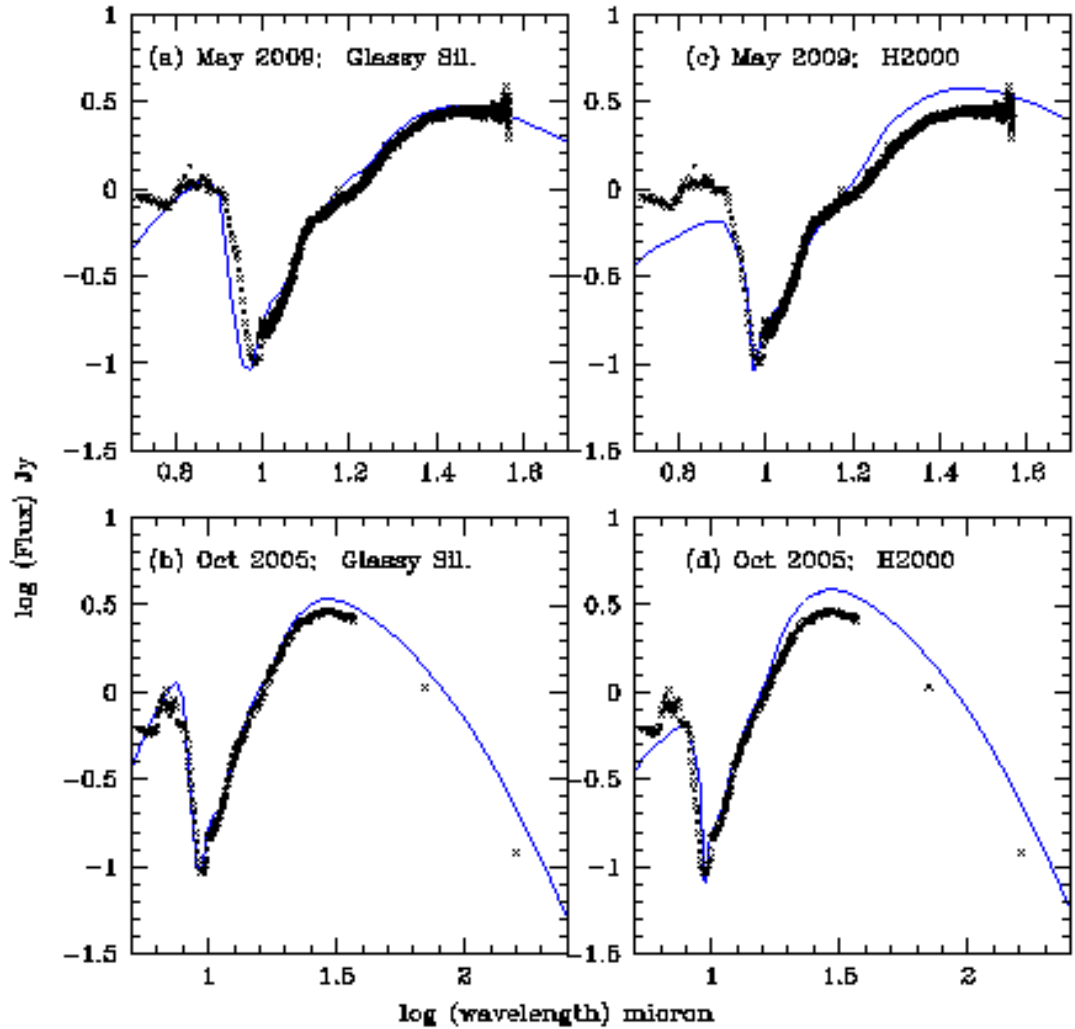

Figure 6. DUSTY fits (in blue) to the observed data of 2005 and 2009 (black crosses) using dust composed of silicates and alumina in different proportions (listed in Table 2). The left side panels show fits using the glassy silicate of Jaeger et al (1994) combined with alumina while the right panel fits are from Hallenbeck et al (2000). For the 2005 data, the 70 and 160 micron Spitzer flux values are also shown. The principal conclusion reached is that the good fits obtained, for both 2005 and 2009 (panels a and b) using the glassy silicate model, indicate a decreasing silicate and increasing alumina content which is consistent with the chaotic silicate hypothesis of dust formation and evolution.